\documentclass[useAMS,usenatbib]{mnras}
\usepackage{graphicx}
\usepackage{amsmath, amssymb}
\usepackage[T1]{fontenc}
\usepackage[usenames]{color}
\usepackage[dvipsnames]{xcolor}
\usepackage{geometry}
\usepackage{bm}
\usepackage[capitalize]{cleveref}
\usepackage{physics}
\usepackage{acro}
\usepackage{caption}
\usepackage{subcaption}
\usepackage{stfloats}
\usepackage{xspace}
\usepackage{booktabs}
\usepackage{tabularx}
\usepackage{multicol}
\usepackage{url}
\usepackage{hyperref}
\usepackage{orcidlink}
\usepackage[group-separator={,}]{siunitx}

\hypersetup{breaklinks=true}

\newcommand{\zobs}{z_{\rm obs}}
\newcommand{\zpred}{z_{\rm pred}}
\newcommand{\zCMB}{z_{\rm CMB}}
\newcommand{\zcosmo}{z_{\rm cosmo}}

\newcommand{\Vpec}{V_{\rm pec}}
\newcommand{\Vext}{\bm{V}_{\rm ext}}

\newcommand{\Mpch}{\ensuremath{h^{-1}\,\mathrm{Mpc}}}
\newcommand{\kmsec}{\ensuremath{\mathrm{km}\,\mathrm{s}^{-1}}}
\newcommand{\kmsecMpc}{\ensuremath{\mathrm{km}\,\mathrm{s}^{-1}\,\mathrm{Mpc}^{-1}}}

\newcommand{\mpred}{m_{\rm pred}}
\newcommand{\mobs}{m_{\rm obs}}
\newcommand{\etaobs}{\eta_{\rm obs}}
\newcommand{\etatrue}{\eta_{\rm true}}

\newcommand{\sint}{\sigma_{\rm int}}

\newcommand{\aTFR}{a_{\rm TFR}}
\newcommand{\bTFR}{b_{\rm TFR}}
\newcommand{\cTFR}{c_{\rm TFR}}

\newcommand{\aFP}{a_{\rm FP}}
\newcommand{\bFP}{b_{\rm FP}}
\newcommand{\cFP}{c_{\rm FP}}

\newcommand{\MSN}{M_{\rm SN}}

\newcommand{\Om}{\Omega_{\rm m}}
\newcommand{\TWOMPP}{2M\texttt{++}\xspace }
\newcommand{\PP}{Pantheon\texttt{+}\xspace }

\defcitealias{Carrick_2015}{C15}

\DeclareAcronym{CMB}{short = CMB, long  = cosmic microwave background}
\DeclareAcronym{LOS}{short = LOS, long  = line-of-sight}
\DeclareAcronym{BORG}{short = \texttt{BORG}, long  = \textit{Bayesian Origin Reconstruction from Galaxies}}
\DeclareAcronym{TFR}{short = TFR, long  = Tully--Fisher relation}
\DeclareAcronym{FP}{short = FP, long  = fundamental plane}
\DeclareAcronym{SN}{short = SN, long  = supernova, short-plural = e, long-plural  = e}
\DeclareAcronym{LCDM}{short = $\Lambda$CDM, long  = $\Lambda$-cold dark matter}
\DeclareAcronym{CF4}{short = CF4, long  = CosmicFlows-4}
\DeclareAcronym{HMC}{short = HMC, long  = Hamiltonian Monte Carlo}
\DeclareAcronym{SDSS}{short = SDSS, long = Sloan Digital Sky Survey}
\DeclareAcronym{WISE}{short = WISE, long = Wide-field Infrared Survey Explorer}
\DeclareAcronym{MCMC}{short = MCMC, long = Markov Chain Monte Carlo}
\DeclareAcronym{NUTS}{short = NUTS, long = No-U-Turn Sampler}

\title[$S_8$ from peculiar velocities]{$S_8$ from peculiar velocities: agreement with \textit{Planck} for Tully--Fisher and supernovae, tension for the fundamental plane}
\author[R. Stiskalek]{Richard Stiskalek$^{1}$\orcidlink{0000-0002-0986-314X}\thanks{\href{mailto:richard.stiskalek@physics.ox.ac.uk}{richard.stiskalek@physics.ox.ac.uk}}
\\
$^{1}$Astrophysics, University of Oxford, Denys Wilkinson Building, Keble Road, Oxford, OX1 3RH, UK\\
}

\date{Accepted XXX. Received YYY; in original form ZZZ}
\pubyear{2025}

\begin{document}\label{firstpage}
\pagerange{\pageref{firstpage}--\pageref{lastpage}}
\maketitle

\begin{abstract}
Peculiar velocity measurements constrain the parameter combination $f\sigma_8$, the product of the linear growth rate $f$ and the fluctuation amplitude $\sigma_8$.
Under the approximation that $f$ is a monotonic function of $\Om$, this can be related to $S_8 \equiv \sigma_8 \sqrt{\Om/0.3}$, enabling direct comparison with weak lensing and cosmic microwave background results.
We use three classes of direct-distance tracers---the Tully--Fisher relation, the fundamental plane, and Type~Ia supernovae---to infer peculiar velocities.
A unified hierarchical forward model jointly calibrates each distance indicator and a linear theory reconstruction of the local Universe.
This is the first consistent Bayesian analysis to treat all three major classes of distance indicators within a common framework, enabling cross-checks of systematics across diverse galaxy populations.
Combining the Tully--Fisher and Type~Ia supernova samples, we obtain $S_8 = 0.798 \pm 0.035$ ($f\sigma_8 = 0.412 \pm 0.018$), in agreement with \textit{Planck} and robust under the choice of galaxy bias model, with the uncertainty dominated by the variance of the \TWOMPP\ galaxy field.
The fundamental plane constraints are instead unstable under the inhomogeneous Malmquist bias treatment; the quadratic extension preferred by the fundamental plane data drives their $S_8$ values lower.
These findings indicate that low-redshift peculiar velocity data are concordant with the cosmic microwave background and do not reinforce the early-versus-late $S_8$ tension, though the fundamental plane results call for further scrutiny of their systematics.
\end{abstract}


%%%%%%%%%%%%%%%%%%%%%%%%%%%%%%%%%%%%%%%%%%%%%%%%%%%%%%%%%%%%%%%%%%%%%%%%%%%%%%%
%                           Introduction                                      %
%%%%%%%%%%%%%%%%%%%%%%%%%%%%%%%%%%%%%%%%%%%%%%%%%%%%%%%%%%%%%%%%%%%%%%%%%%%%%%%

\section{Introduction}\label{sec:intro}

The standard cosmological model, \ac{LCDM}, reproduces the large-scale properties of the Universe, from the anisotropies of the \ac{CMB} to the distribution of galaxies at low redshift~\citep{Peebles1980,Peebles_2003}.
It is built on the assumptions of statistical homogeneity and isotropy, which imply the Friedmann--Robertson--Walker metric, and on general relativity to describe the dynamics of space-time.
The dominant energy components are cold dark matter and a cosmological constant $\Lambda$, which provide a minimal framework that explains a wide range of observations with only a handful of parameters.

Despite its successes, \ac{LCDM} must be tested with independent observables that probe different physical regimes~\citep{Perivolaropoulos_2022,Secrest_2022,cosmoverse}.
One of its predictions is the rate at which density perturbations grow under gravity, commonly referred to as the growth rate of structure.
In the linear regime, the growth rate follows a simple dependence on the matter density, $f \approx \Om^{0.55}$~\citep{Bouchet_1995,Wang_1998}.
Departures from this relation would indicate physics beyond the standard model, such as modifications to gravity or dark energy (e.g.~\citealt{Dvali_2000, Linder_2007}).

The growth rate is sometimes measured in combination with the parameter $\sigma_8$, which describes the amplitude of density fluctuations on $8~\Mpch$ scales, in the form of $f\sigma_8$.
Weak lensing surveys are primarily sensitive to the degenerate combination of $\sigma_8$ and $\Om$, and constraints are therefore commonly expressed through $S_8 \equiv \sigma_8 \sqrt{\Om/0.3}$~\citep{Mandelbaum_2018}.
Over the past decade these surveys have systematically reported lower values of $S_8$ than those inferred from the cosmic microwave background by the \textit{Planck} experiment~\citep{Planck_2020_cosmo}, giving rise to the $S_8$ tension~\citep{Heymans_2013,Hikage_2019,Asgari_2021,DES_Y3,DES_KIDS,Carlos_2024,Wright_2025,Gomes_2025}.
Independent low-redshift probes of $S_8$ are therefore essential to establish whether this discrepancy reflects new physics or residual systematics.

Measurements of peculiar velocities provide a low-redshift test of the cosmological model.
The velocity field of the local Universe is sourced by the gravitational potential of the \emph{total} matter distribution.
This relation is given by the continuity equation, which in the linear regime relates the divergence of the velocity field to the matter overdensity~\citep{Weinberg_2008}.
On smaller scales, non-linear motions can be resolved e.g.~with $N$-body simulations~\citep{Angulo_2022}.
A key difference between peculiar velocities and clustering is that peculiar velocities are sensitive to large-scale modes: the velocity power spectrum is proportional to the matter power spectrum divided by $k^2$, where $k$ denotes the Fourier wavenumber~\citep{Coles_2002}.

Peculiar velocities are not observable, since galaxy distances cannot be measured directly and must instead be inferred from distance-indicating observables.
In practice, the distance should be treated as a latent parameter and inferred statistically, most naturally within a Bayesian framework (e.g.~\citealt{Trotta_2008}).
A variety of distance indicators exist, typically relating a distance-independent observable to either the intrinsic luminosity or the physical size of the source.
In this work we employ three complementary classes: the \ac{TFR} for late-type galaxies~\citep{Tully_1977}, the \ac{FP} relation for early-type galaxies~\citep{Djorgovski_1987,Dressler_1987}, and Type~Ia \acp{SN} as standardisable candles~\citep{Phillips_1993,Riess_1996}.
Together, these probe diverse galaxy populations with distinct systematics.
We combine all three tracers with the linear theory density and velocity field of the local Universe modelled by~\citet[hereafter \citetalias{Carrick_2015}]{Carrick_2015} based on the \TWOMPP\ galaxy sample~\citep{Lavaux_2011}.
We analyse all datasets within a single Bayesian framework that jointly calibrates the distance indicators and rescales the velocity field to match the observed redshifts.

Redshift-space distortions are well established as probes of the growth rate (e.g.~\citealt{Beutler_2012,Adams_2020,Blake_2021,Turner_2021,Turner_2023}).
However, no previous analysis has jointly incorporated the \ac{TFR}, \ac{FP}, and \acp{SN}.
Similar measurements of $S_8$ (or $f\sigma_8$) have relied on the \ac{TFR} alone~\citep{Carrick_2015,Boubel_2024}, \acp{SN} alone~\citep{Stahl_2021}, on combined \ac{TFR} and \acp{SN} samples~\citep{Boruah_2019,VFO}, or only on the \ac{FP}~\citep{Said_2020}.
Moreover, these studies adopt heterogeneous methodologies, which hampers a clean comparison between results.
Peculiar velocity (or ``direct-distance'') surveys such as CosmicFlows~\citep{CosmicFlows2,Tully_2016,Tully_2023}, SFI\texttt{++}~\citep{Masters_2006, Springob_2007}, 2MTF~\citep{Masters_2008,Hong_2019}, 6dF~\citep{Springob_2014}, and \ac{SDSS} \ac{FP}~\citep{Howlett_2022} have mapped local flows and constrained cosmology for decades.
Building on our previous work~\citep{VFO,CH0,dH0}, we extend this approach to an analysis of all three tracers.

The paper is structured as follows.
In~\cref{sec:local_universe} we describe the linear theory reconstruction of the local Universe.
In~\cref{sec:pv_samples} we introduce the peculiar velocity samples used in this work, in~\cref{sec:method} we outline the methodology for jointly calibrating the distance-indicator relation and the velocity field, and in~\cref{sec:results} we present our measurement of $S_8$.
In~\cref{sec:comparison_to_literature} we compare our measurements with literature estimates, and in~\cref{sec:conclusion} we conclude.
All logarithms are base-10 unless explicitly stated.
We denote by $\mathcal{N}(x; \mu, \sigma)$ the normal distribution with mean $\mu$ and standard deviation $\sigma$ evaluated at $x$.
Throughout, distances are expressed in $\Mpch$, with $h \equiv H_0 / (100~\kmsecMpc)$.

%%%%%%%%%%%%%%%%%%%%%%%%%%%%%%%%%%%%%%%%%%%%%%%%%%%%%%%%%%%%%%%%%%%%%%%%%%%%%%%
%                               Data                                          %
%%%%%%%%%%%%%%%%%%%%%%%%%%%%%%%%%%%%%%%%%%%%%%%%%%%%%%%%%%%%%%%%%%%%%%%%%%%%%%%

\section{Local Universe model}\label{sec:local_universe}

We adopt the reconstruction of~\citetalias{Carrick_2015} as a model of the density and peculiar velocity field of the local Universe derived from the \TWOMPP\ catalogue~\citep{Lavaux_2011}.
\TWOMPP\ is a whole-sky redshift-space compilation of \num{69160} galaxies, derived using photometry from the Two-Micron-All-Sky Extended Source Catalog~\citep{Skrutskie_2006} and redshifts from the 2MASS Redshifts Survey~\citep[2MRS,][]{Huchra_2012}, the 6dF Galaxy Redshift Survey~\citep{Jones_2009}, and the \ac{SDSS} Data Release 7~\citep{Abazajian_2009}.
The \TWOMPP\ $K$-band apparent magnitudes are corrected for Galactic extinction, $k$-corrections, and surface-brightness dimming.
The catalogue is complete to $K < 11.5$ in the regions covered by 2MRS and to $K < 12.5$ within the 6dF and \ac{SDSS} footprints, corresponding to a depth of ${\sim} 200$~Mpc for galaxies with luminosities near the knee of the luminosity function.

\citetalias{Carrick_2015} employ luminosity-based galaxy weighting to account for sample completeness, assuming a constant linear bias $b_{\mathrm{2M\texttt{++}}}$.
The redshift-space galaxy distribution is voxelised and smoothed with a Gaussian kernel of standard deviation $4~\Mpch$.
From this density field, the velocity field can be obtained under the assumption of \ac{LCDM} linear theory~\citep{Peebles1980}, though the density field itself does not assume \ac{LCDM}:
\begin{equation}\label{eq:density_to_velocity}
    \bm{v}(\bm{r}) = \frac{H_0 f}{4\pi} \int \mathrm{d}^3 \bm{r}' \, \delta(\bm{r}') \frac{\bm{r}' - \bm{r}}{|\bm{r}' - \bm{r}|^3},
\end{equation}
where $\delta(\bm{r})$ is the density contrast at position $\bm{r}$.
When distances are expressed in units of $\Mpch$, the $H_0$ dependence in~\cref{eq:density_to_velocity} cancels.
The linear growth rate is defined as $f \equiv \mathrm{d} \ln D / \mathrm{d} \ln a$, with $D$ the growth function of linear perturbations and $a$ the scale factor.
In \ac{LCDM}, $f \approx \Om^{0.55}$~\citep{Bouchet_1995,Wang_1998}, though in modified gravity theories this index differs from $0.55$ (e.g.~\citealt{Dvali_2000,Linder_2007}).
The redshift-space galaxy distribution can then be mapped back to real space using the iterative scheme of~\citet{Yahil_1991}.

Assuming the galaxy field is related to the matter field as
\begin{equation}\label{eq:matter_to_galaxy_bias}
    \delta_g(\bm{r}) = b_{\mathrm{2M\texttt{++}}} \, \delta(\bm{r}),
\end{equation}
where $b_{\mathrm{2M\texttt{++}}}$ is the linear bias of the \TWOMPP\ sample relative to the matter field\footnote{Not to be confused with the inhomogeneous Malmquist bias parameter $b_1$ introduced in~\cref{eq:linear_bias}, which describes the bias of the peculiar velocity tracer relative to the \TWOMPP\ density field.}, we may replace the matter field in~\cref{eq:density_to_velocity} with the luminosity-weighted galaxy density contrast.
This introduces the degenerate combination $\beta^\star \equiv f / b_{\mathrm{2M\texttt{++}}}$ in front of the integral.
From~\cref{eq:matter_to_galaxy_bias} it follows that $\sigma_8^g =  b_{\mathrm{2M\texttt{++}}} \, \sigma_8^{\rm NL}$, where $\sigma_8^g$ is the fluctuation amplitude of the galaxy density field on $8\,\Mpch$ scales, and $\sigma_8^{\rm NL}$ is the corresponding fluctuation amplitude of the non-linear matter field.
$\sigma_8^g$ has been measured for the \TWOMPP\ sample by~\citet{westover,Carrick_2015}.
$\beta^\star$ can be inferred from peculiar velocity data, since peculiar velocities are sourced by the total matter distribution (e.g.~\citealt{Carrick_2015,Boruah_2019,Said_2020,Boubel_2024,VFO}).
Thus, $\beta^\star$ inferred from peculiar velocities, together with $\sigma_8^g$ measured from the clustering of the \TWOMPP\ sample, constrains the parameter combination
\begin{equation}
    f \, \sigma_8^{\rm NL} = \beta^\star \, \sigma_8^g,
\end{equation}
which can be related to
\begin{equation}
    S_8 \equiv \sigma_{8}^{\rm L} \sqrt{\Om / 0.3}.
\end{equation}
Since we assume $f \approx \Om^{0.55}$, the explicit dependence on $\Om$ largely cancels in the definition of $S_8$.
The remaining step is to relate the fluctuation amplitude of the non-linear matter field, $\sigma_8^{\rm NL}$, to the fluctuation amplitude of the linear matter field $\sigma_8^{\rm L}$ used in the definition of $S_8$.
A mapping between the two was derived by~\citet{Juszkiewicz_2010}, but we showed in~\citet{VFO} that their approximation is inaccurate at the level of $3$--$5$ per cent.
Instead, we computed non-linear matter power spectra using the \texttt{syren-new} emulator~\citep{Bartlett_2024,Sui_2024}.
For each model we varied the primordial power spectrum amplitude $A_{\rm s}$, using a root-finding algorithm to determine the value that reproduces a given $\sigma_8^{\rm NL}$ from the integral of the non-linear power spectrum.
This $A_{\rm s}$ was then converted to the corresponding linear variance $\sigma_8^{\rm L}$ using the prescription of~\citet{Sui_2024}.
We assumed a flat \ac{LCDM} cosmology with $h = 0.6766$, $\Om = 0.3111$, $\Omega_{\rm b} = 0.02242 / h^2$, and $n_s = 0.9665$, though our results are not sensitive to this choice.

The density and velocity fields of~\citetalias{Carrick_2015} are generated on a $256^3$ grid with a box size of $400~\Mpch$.
Using a maximum-likelihood counts-in-cells scheme within radial shells~\citep{Efstathiou_1990}, \citetalias{Carrick_2015} measured $\sigma_8^g = 0.99 \pm 0.04$.

\section{Peculiar velocity samples}\label{sec:pv_samples}

We use three direct-distance tracers: the \ac{TFR}, \ac{FP}, and Type~Ia \acp{SN}, whose samples we describe below.

\subsection{Tully--Fisher relation}\label{sec:TFR_data}

The \ac{TFR}~\citep{Tully_1977} is an empirical scaling relation linking the rotation velocity of spiral galaxies, traced by the width of a spectral line $W$ (most commonly the H\textsc{I} line), to their absolute magnitude $M$, which serves as a proxy for luminosity. To reparameterise the linewidth $W$, we define
\begin{equation}
    \eta \equiv \log \frac{W}{\kmsec} - 2.5,
\end{equation}
and henceforth refer to $\eta$ simply as the linewidth. We express the \ac{TFR} as
\begin{equation}\label{eq:TFR_absmag}
    M(\eta) =
    \begin{cases}
        \aTFR + \bTFR \eta + \cTFR \eta^2, & \eta > 0, \\
        \aTFR + \bTFR \eta, & \eta \leq 0,
    \end{cases}
\end{equation}
where $\aTFR$, $\bTFR$, and $\cTFR$ denote the zero-point, slope, and curvature, respectively.
We account for curvature of the \ac{TFR} when calibrating high-linewidth galaxies ($\eta > 0$).
We infer the calibration parameters of the \ac{TFR} jointly with its intrinsic scatter, $\sint$.

In this work, we use the state-of-the-art \ac{CF4} \ac{TFR} survey, which comprises \num{9792} galaxies with $z \lesssim 0.05$~\citep{Kourkchi_2020B,Kourkchi_2020A}, and forms part of the broader \ac{CF4} sample~\citep{Tully_2023}.
We make use of both the \ac{SDSS} $i$ and \ac{WISE} W1 photometry: the $i$ band is restricted to the \ac{SDSS} footprint, whereas the W1 band covers the full sky.
Although some galaxies have measurements in both bands, we treat them as separate samples; except when combining them in a single inference, we preferentially use \ac{SDSS} photometry.
From this parent sample we select galaxies with $\eta > -0.3$, Galactic latitude $|b| > 7.5^\circ$, and quality flag $5$ (best).
The resulting \ac{SDSS} $i$-band and \ac{WISE} W1 samples contain \num{5027} and \num{3278} galaxies, respectively.

\subsection{Fundamental plane relation}\label{sec:FP_relation}

The \ac{FP} relation provides a distance indicator for early-type galaxies, linking their effective radius, velocity dispersion, and mean surface brightness~\citep{Djorgovski_1987,Dressler_1987}. In logarithmic form, the \ac{FP} is expressed as
\begin{equation}\label{eq:FP_relation}
    \log R_e = \aFP \log \sigma_0 + \bFP \log I_e + \cFP,
\end{equation}
where $R_e$ is the effective radius in physical units, $\sigma_0$ the central velocity dispersion (aperture corrected;~\citealt{Jorgensen_1995}), and $I_e$ the mean surface brightness within $R_e$.
The three \ac{FP} coefficients are $\aFP,\,\bFP$, and $\cFP$.
The right-hand side of~\cref{eq:FP_relation} contains only distance-independent observables, while
\begin{equation}
    R_e \approx \theta_{\rm eff}\, d_A,
\end{equation}
with $\theta_{\rm eff}$ the observed angular size and $d_A$ the angular-diameter distance.
This relation therefore enables redshift-independent distance measurements.
We shall infer the three \ac{FP} parameters, along with the \ac{FP} intrinsic scatter in $\log \theta_{\rm eff}$.
We employ two \ac{FP} samples: the \ac{SDSS} \ac{FP} catalogue by~\citet{Howlett_2022} and the 6dF \ac{FP} catalogue by~\citet{Campbell_2014}.

The \ac{SDSS} \ac{FP} catalogue contains \num{34059} early-type galaxies selected from \ac{SDSS} DR14~\citep{Abolfathi_2018}, making it the largest \ac{FP} sample to date.
Galaxies are selected to be red, quiescent, dispersion-supported systems, and the final sample has a mean fractional distance error of 23 per cent.
The 6dF \ac{FP} catalogue contains nearly \num{11000} early-type galaxies across the southern sky, drawn from the brightest ellipticals, lenticulars, and early-type spiral bulges.
For consistency with~\citetalias{Carrick_2015} and our other samples, we restrict both catalogues to $\zCMB < 0.05$, yielding \num{7447} \ac{SDSS} and \num{8722} 6dF galaxies.

\subsection{Type Ia supernovae}\label{sec:SN_data}

Type~Ia \acp{SN} serve as ``standardisable'' candles in cosmology. Their light curves are standardised with the SALT2 model~\citep{SALT2}, yielding a standardised apparent magnitude through the Tripp formula~\citep{Tripp_1998}:
\begin{equation}\label{eq:tripp_formula}
    m_{\rm standard} = \mobs + \mathcal{A} x_1 - \mathcal{B} c,
\end{equation}
where $\mobs$ is the observed \ac{SN} apparent magnitude, $x_1$ the light-curve stretch, and $c$ the colour.
The global parameters $\mathcal{A}$ and $\mathcal{B}$ quantify the stretch and colour corrections, respectively.
Combined with the absolute magnitude $\MSN$, the standardised magnitude $m_{\rm standard}$ yields the distance modulus.

We use the \PP\ compilation, which contains \num{1701} spectroscopically confirmed Type~Ia \acp{SN} spanning redshifts from $z {\sim} 0.001$ to ${\sim} 2.3$~\citep{Scolnic_2022,Brout_2022,Peterson_2022,Carr_2022}.
However, to match the redshift range of~\citetalias{Carrick_2015} we restrict to $\zCMB \leq 0.05$, resulting in a subset of \num{525} \acp{SN}.
In \PP, distances are derived with the SALT2 fitter and corrected for selection effects using the BEAMS with Bias Corrections (BBC) method~\citep{Kessler_2017}, which introduces an additive bias term to the magnitudes in~\cref{eq:tripp_formula}.
We adopt these bias-corrected magnitudes, $m_{\rm corr}$, which also include a fiducial Tripp calibration.
Consequently, we infer only the standardised absolute magnitude $\MSN$, while keeping the stretch and colour coefficients fixed to their assumed fiducial values.

Uncertainties in the standardised magnitudes (or equivalently, the distance moduli) are provided through a covariance matrix that incorporates both statistical and systematic contributions, including the uncertainties in $\mathcal{A}$ and $\mathcal{B}$ held fixed at their fiducial values.
While the full covariance matrix provided in the \PP\ release includes contributions from peculiar velocities, our model accounts for these explicitly.
We therefore use a reduced version of the covariance matrix with the peculiar velocity terms removed, as provided to us by Anthony Carr (priv. comm.).

%%%%%%%%%%%%%%%%%%%%%%%%%%%%%%%%%%%%%%%%%%%%%%%%%%%%%%%%%%%%%%%%%%%%%%%%%%%%%%%
%                               Methodology                                   %
%%%%%%%%%%%%%%%%%%%%%%%%%%%%%%%%%%%%%%%%%%%%%%%%%%%%%%%%%%%%%%%%%%%%%%%%%%%%%%%

\section{Joint distance and velocity calibration}\label{sec:method}

\begin{figure*}
    \centering
    \includegraphics[width=1\textwidth]{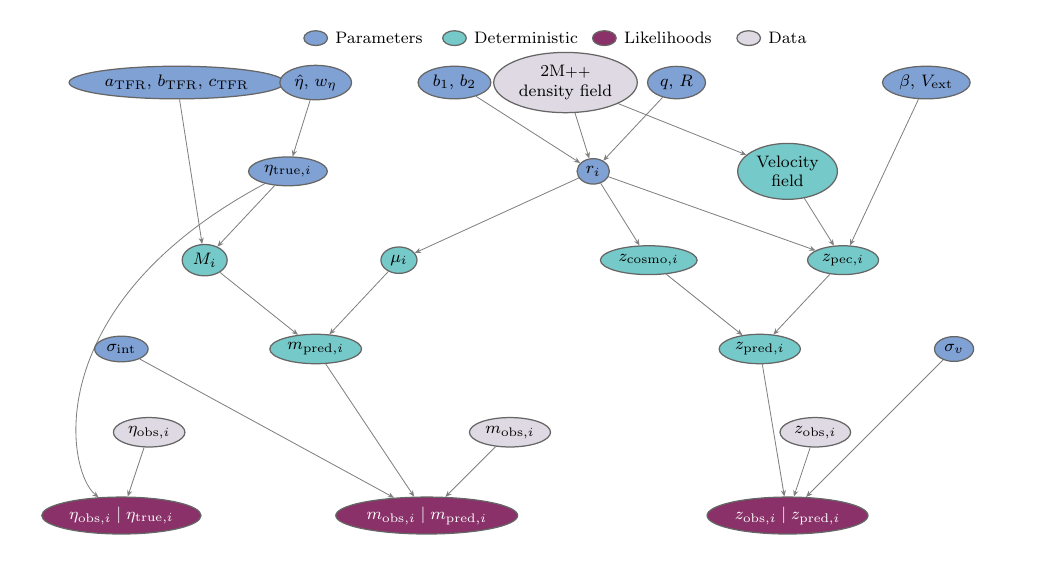}
    \caption{Directed acyclic graph of our \ac{TFR} model. An analogous model applies to the \ac{FP} and \acp{SN}.
    The density and velocity fields are external data products from~\protect\citet{Carrick_2015}, whose velocity field is a deterministic function of their density field.
    }
    \label{fig:TFR_DAG}
\end{figure*}

We adopt a Bayesian forward-modelling approach, constructing a hierarchical model that maps the parameters $\bm\theta$ to the observables while marginalising over all nuisance parameters.
Here $\bm\theta$ denotes the full parameter space, including the distance-relation parameters (e.g.~\ac{TFR} or \ac{FP} coefficients), velocity scaling $\beta^\star$, velocity scatter $\sigma_v$, and external velocity dipole $\Vext$.
The model is adapted to each class of distance indicator (\ac{TFR}, \ac{FP}, or Type~Ia \acp{SN}).
This framework was introduced in our previous work for \ac{TFR} and \acp{SN} samples~\citep{VFO,dH0}, and here we extend it to include the \ac{FP}.

\begin{figure}
    \centering
    \includegraphics[width=\columnwidth]{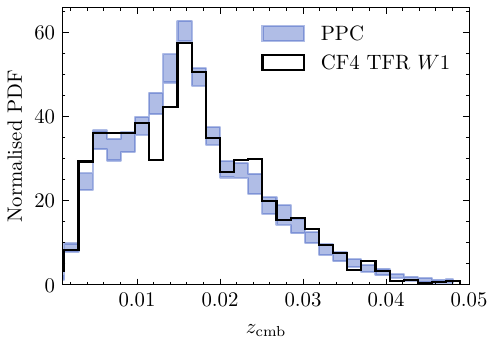}
    \caption{Posterior predictive check (PPC) for the \ac{CF4} \ac{TFR} W1 sample. The black histogram shows the observed \ac{CMB}-frame redshift distribution; the shaded band shows the $1\sigma$ scatter of draws from the forward model using the mean global posterior parameters listed in~\cref{tab:b1_beta_S8}.}
    \label{fig:ppc_W1_zcmb}
\end{figure}

Since peculiar velocity samples are subject to complex and poorly understood selection in distance-correlated quantities (such as optical magnitude or H\textsc{I} flux), we model this effect following the phenomenological prior of~\citet{Lavaux_Virbius}.
We place the prior on the 3D source position $\bm{r}$,
\begin{equation}\label{eq:empirical_prior_3D}
    \pi(\bm{r} \mid \bm{\theta})
    = \frac{n(\bm{r},\, \bm{\theta})\, f(|\bm{r}|,\, \bm{\theta})}
           {Z(\bm{\theta})},
\end{equation}
with the normalisation
\begin{equation}\label{eq:empirical_prior_3D_norm}
    Z(\bm{\theta}) \equiv \int_V \mathrm{d}^3\bm{r}'\, n(\bm{r}',\, \bm{\theta})\, f(|\bm{r}'|,\, \bm{\theta}),
\end{equation}
where $V$ denotes the reconstruction volume of~\citetalias{Carrick_2015}, $n(\bm{r},\,\bm{\theta})$ encodes the inhomogeneous Malmquist bias from the~\citetalias{Carrick_2015} luminosity density evaluated at $\bm{r}$, and
\begin{equation}\label{eq:fr_completeness}
    f(r,\, \bm{\theta}) = \exp\!\left[-\left(\frac{r}{R}\right)^q\right]
\end{equation}
parametrises the radial sample incompleteness through a characteristic depth $R$ and a sharpness $q$, both inferred jointly with the model.
The angular position of each source is measured with negligible uncertainty, so its likelihood is effectively a delta function fixing the latent direction $\hat{\bm{r}}$ to its observed value $\hat{\bm{r}}_{\rm obs}$.
Writing $\bm{r} = r\,\hat{\bm{r}}_{\rm obs}$ and changing to spherical coordinates yields the radial prior
\begin{equation}\label{eq:empirical_prior_distance}
    \pi(r \mid \hat{\bm{r}}_{\rm obs},\,\bm{\theta})
    = \frac{r^2\,n(r\,\hat{\bm{r}}_{\rm obs},\,\bm{\theta})\,f(r,\,\bm{\theta})}{Z(\bm{\theta})},
\end{equation}
where the Jacobian factor $r^2$ is fixed by the volume element.
As a posterior predictive check,~\cref{fig:ppc_W1_zcmb} compares the observed \ac{CMB}-frame redshift distribution of the \ac{CF4} \ac{TFR} W1 sample to forward draws from this empirical distance prior combined with the inhomogeneous Malmquist bias, using the mean global posterior parameters that we shall infer for this sample (\cref{tab:b1_beta_S8}).
The two distributions agree closely.
A more rigorous treatment of selection effects, which requires knowledge of the survey selection, is discussed e.g.~by~\citet{Kelly_2008} (and we recently applied it in~\citealt{CH0}).

We model the inhomogeneous Malmquist bias term as
\begin{equation}\label{eq:linear_bias}
    n(\bm{r},\,b_1) = 1 + b_1 \delta(\bm{r}),
\end{equation}
where $\delta(\bm{r})$ is the luminosity density contrast from~\citetalias{Carrick_2015} at the galaxy's position.
Since $\delta(\bm{r})$ traces the luminosity of the \TWOMPP\ galaxy field with bias $b_{\mathrm{2M\texttt{++}}}$ relative to the matter field, the parameter $b_1$ here represents the ratio of the linear bias of the peculiar velocity tracer to that of \TWOMPP.
However, this formulation implies non-physical values $n(\bm{r},\,b_1) < 0$ when $\delta < - 1 / b_1$.
To avoid this, we use a quadratic smoothing
\begin{equation}\label{eq:smoothmax_bias}
    n \rightarrow \frac{1}{2}\!\left(n + \sqrt{n^2 + \tau^2}\right),
\end{equation}
where $n \equiv n(\bm{r},\,b_1)$ and we set $\tau = 0.01$, though the presented results do not depend on the exact value (even if we had clipped $n$ at zero).
This linear bias model is consistent with~\citetalias{Carrick_2015},
who relate the $4~\Mpch$-smoothed galaxy field to the matter density via a linear bias, and the matter density to the velocity field via linear theory.
We adopt a uniform prior on $b_1$ and introduce a different $b_1$ for each peculiar velocity sample.
As a robustness check we also consider a quadratic extension,
\begin{equation}\label{eq:quadratic_bias}
    n(\bm{r},\,b_1,\,b_2) = 1 + b_1 \delta(\bm{r}) + b_2 \delta^2(\bm{r}),
\end{equation}
with a standard normal prior $\pi(b_2) = \mathcal{N}(b_2;\,0,\,1)$, admitting the negative values found in $N$-body simulations for haloes of mass $\log_{10}(M/h^{-1}M_\odot) \lesssim 13.5$, the regime hosting our peculiar velocity tracers~\citep{Lazeyras_2016}.
The likelihood of the observed redshift $\zobs$ (converted to the \ac{CMB} frame) is
\begin{equation}\label{eq:redshift_likelihood}
    \mathcal{L}(\zobs \mid r,\,\bm{\theta}) =
    \mathcal{N}\!\left(\zobs;\,\zpred,\,\sqrt{\sigma_v^2 + \sigma_{cz}^2}\right),
\end{equation}
where $\zpred$ is the predicted redshift, which depends on the galaxy distance $r$ and the \ac{LOS} peculiar velocity $\Vpec$ as
\begin{equation}
    1 + \zpred = (1 + \zcosmo)\,(1 + \Vpec / c),
\end{equation}
with $\zcosmo$ being the cosmological redshift at distance $r$. (We work with distance in units of $\Mpch$, and since the zero-point calibration is degenerate with $\log h$, we sample their degenerate parameter combination.) The \ac{LOS} peculiar velocity is given by
\begin{equation}
    \Vpec(\bm{r}) = \left(\beta^\star \bm{v}(\bm{r}) + \Vext\right) \cdot \hat{\bm{r}}_{\rm obs},
\end{equation}
where $\bm{v}(\bm{r})$ is the velocity from~\citetalias{Carrick_2015}, which is computed according to~\cref{eq:density_to_velocity} with $\beta^\star = 1$, so that the scaling by $\beta^\star$ is applied only at this stage. The term $\Vext$ represents a constant external velocity dipole, the leading-order multipole accounting for flows sourced by matter outside the reconstruction volume, and $\hat{\bm{r}}_{\rm obs}$ is the unit vector along the source \ac{LOS}.
In~\cref{eq:redshift_likelihood} we assume that, once peculiar velocities are modelled using~\citetalias{Carrick_2015}, the residuals are uncorrelated and well described by a Gaussian distribution whose standard deviation is given by $\sigma_v$ (the residual velocity scatter) and the spectroscopic redshift uncertainty $\sigma_{cz}$ (negligible relative to $\sigma_v$). We adopt a uniform prior for $\beta^\star$, a Maxwell--Boltzmann prior with mean $\langle\sigma_v\rangle \approx 200~\kmsec$ for the residual velocity scatter, motivated by interpreting $\sigma_v$ as the magnitude of an isotropic 3D Gaussian residual velocity\footnote{Explicitly, $\pi(\sigma_v) \propto \sigma_v^2 \exp[-\sigma_v^2 / (2 a^2)]$ with scale $a = 125~\kmsec$, giving $\langle\sigma_v\rangle = 2a\sqrt{2/\pi} \approx 200~\kmsec$.} (we have verified that this prior choice has a negligible effect on the inference), and a prior on $\Vext$ that is uniform in both magnitude and sky direction.

For the \ac{TFR}, we place a Gaussian hyperprior on $\etatrue$ with mean $\hat{\eta}$ (inferred under a uniform prior) and width $w_\eta$ (inferred with a reference prior $\pi(w_\eta) \propto 1 / w_\eta$). These hyperparameters are inferred jointly with all other model parameters in a hierarchical framework. The choice of a Gaussian hyperprior is motivated by the approximate Gaussianity of the observed linewidth distributions and follows similar approaches in \ac{SN} cosmology~\citep{March_2011,March_2014,Rubin_2015,March_2018,Rubin_2023}. The measurement error on $\etaobs$ is assumed Gaussian with standard deviation $\sigma_\eta$, giving the likelihood
\begin{equation}\label{eq:eta_obs_likelihood}
    \mathcal{L}(\etaobs \mid \etatrue) = \mathcal{N}(\etaobs;\,\etatrue,\,\sigma_\eta).
\end{equation}
The implications of truncation in $\etaobs$ are discussed in our earlier work~\citep{VFO}, which we apply consistently here, though the effect on the results is negligible.
In brief, following~\citet{Kelly_2008}, we introduce a term $p(S = 1 \mid \etaobs)$ that describes the fraction of retained samples after selection in $\etaobs$.
This term modifies the model probability density (the product of the individual-sample likelihoods and the prior) by a factor $\left[p(S = 1 \mid \etaobs)\right]^{-n}$, where $n$ is the number of observed sources.
We model this selection jointly with the phenomenological treatment of distance selection through~\cref{eq:empirical_prior_distance}.

Given $\etatrue$, we predict the galaxy absolute magnitude $M(\etatrue)$ following~\cref{eq:TFR_absmag}.
Combined with the source distance, this yields the predicted apparent magnitude
\begin{equation}
    \mpred = \mu(r) + M(\etatrue),
\end{equation}
where $\mu(r)$ is the distance modulus at distance $r$.
We assume that both the intrinsic scatter of the \ac{TFR}, $\sint$, and the (subdominant) measurement error in magnitude, $\sigma_m$, are Gaussian.
The likelihood of the observed magnitude is therefore
\begin{equation}\label{eq:TFR_magnitude_likelihood}
    \mathcal{L}(\mobs \mid \mpred) = \mathcal{N}\!\left(\mobs;\,\mpred,\,\sqrt{\sint^2 + \sigma_m^2}\right).
\end{equation}
We assume uniform priors on $\aTFR,\,\bTFR$, and $\cTFR$ (\ac{TFR} zero-point, slope, and curvature) and a reference prior on $\sint$: $\pi(\sint)\propto 1/ \sint$.
Since we calibrate the distance relation and account for peculiar velocities with~\citetalias{Carrick_2015} and $\Vext$, we treat the sources as mutually independent.
This assumption may affect inferred uncertainties~\citep{Blake_2024}, as discussed in~\cref{sec:conclusion}.
Following the approach introduced in~\citet{VFO}, we numerically marginalise over both $r$ and $\etatrue$ for each galaxy at every \ac{MCMC} step.
\Cref{fig:TFR_DAG} shows the directed acyclic graph of the \ac{TFR} model.

For the \ac{FP} samples, we adopt a similar treatment.
We introduce the true velocity dispersion $\log \sigma_{0,\mathrm{true}}$ and surface brightness $\log I_{e,\mathrm{true}}$, assigning them a two-dimensional correlated Gaussian hyperprior with means inferred under uniform priors, standard deviations under reference priors, and correlation coefficient inferred with a uniform prior between $-1$ and $1$.
This leads to Gaussian likelihoods for the observed velocity dispersion and surface brightness, analogous to~\cref{eq:eta_obs_likelihood}.
The key difference is that the formulation of the \ac{FP} naturally lends itself to predicting the logarithm of the effective angular size,
\begin{equation}
\begin{split}
    \log &\theta_{\rm eff,\,pred} \\
    &= \left(\aFP \log \sigma_{0, \mathrm{true}}
       + \bFP \log I_{e, \mathrm{true}}
       + \cFP \right) - \log d_A,
\end{split}
\end{equation}
where $d_A$ is the angular-diameter distance at the source distance $r$.
The likelihood of the observed effective angular size is then
\begin{equation}
\begin{split}
    \mathcal{L}(&\log \theta_{\rm eff,\,obs} \mid \log \theta_{\rm eff,\,pred})\\
    &= \mathcal{N}\!\left(\log \theta_{\rm eff,\,obs};\, \log \theta_{\rm eff,\,pred},\, \sqrt{\sint^2 + \sigma_{\log \theta_{\rm eff}}^2}\right),
\end{split}
\end{equation}
with $\sint$ being the \ac{FP} intrinsic scatter and $\sigma_{\log \theta_{\rm eff}}$ the measurement error on $\theta_{\rm eff}$ converted to logarithmic units. As with the \ac{TFR}, we treat the galaxies as independent and numerically marginalise over $r$ at every \ac{MCMC} step.
We additionally marginalise analytically over $\log \sigma_{0,\mathrm{true}}$ and $\log I_{e,\mathrm{true}}$, since they enter the Gaussian likelihood linearly and have a Gaussian hyperprior.

For \PP, the standardised magnitudes make the treatment more straightforward.
We predict the standardised magnitude as
\begin{equation}
    \mpred = \mu(r) + \MSN,
\end{equation}
where $\MSN$ is the standardised absolute magnitude, inferred with a uniform prior.
The key difference is that since the magnitudes are standardised, their covariance must be explicitly accounted for in the likelihood.
Denoting by $\bm{m}_{\rm standard}$ the vector of \PP\ standardised apparent magnitudes and by $\bm{m}_{\rm pred}$ the corresponding predicted values, the likelihood is
\begin{equation}
    \mathcal{L}(\bm{m}_{\rm standard} \mid \bm{m}_{\rm pred})
    =
    \mathcal{N}\!\left(\bm{m}_{\rm standard};\,\bm{m}_{\rm pred},\,\mathbf{C}_{\rm SN}\right),
\end{equation}
where $\mathbf{C}_{\rm SN}$ is the \ac{SN} covariance matrix.
Because of this covariance, the host galaxy distances are correlated and no longer reducible to a series of one-dimensional integrals.
We therefore instead sample the distances $r$ explicitly.
For \PP, this results in a posterior with over 500 dimensions (one distance per \ac{SN}, plus global calibration and flow parameters).

We implement all models in \texttt{JAX}\footnote{\url{https://github.com/jax-ml/jax}}, which leverages automatic differentiation to compute gradients. To sample the posterior distribution, we use the \texttt{numpyro}\footnote{\url{https://num.pyro.ai/en/latest/}} package~\citep{Phan_2019, Bingham_2019}, specifically the \acl{NUTS} method of \acl{HMC} sampling~\citep{Hoffman_2011}. We run four independent chains of \num{5000} samples each, discarding the first \num{1000} as burn-in. Convergence is ensured by requiring the Gelman--Rubin statistic $\hat{R}-1 \leq 0.01$ for all parameters~\citep{Gelman_1992}.

%%%%%%%%%%%%%%%%%%%%%%%%%%%%%%%%%%%%%%%%%%%%%%%%%%%%%%%%%%%%%%%%%%%%%%%%%%%%%%%%
%                              Results                                         %
%%%%%%%%%%%%%%%%%%%%%%%%%%%%%%%%%%%%%%%%%%%%%%%%%%%%%%%%%%%%%%%%%%%%%%%%%%%%%%%%

\section{Results}\label{sec:results}

\begin{figure*}
    \centering
    \includegraphics[width=\textwidth]{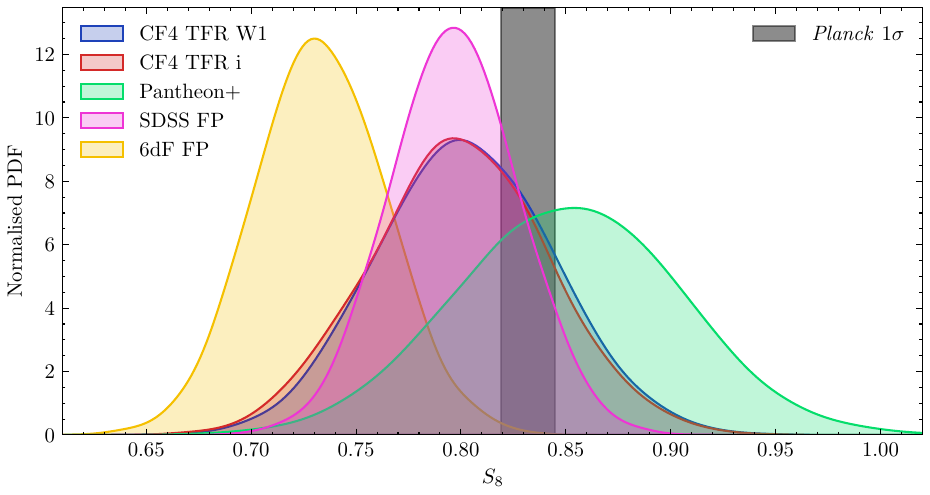}
    \caption{Posterior predictive distributions of $S_8$ under the linear galaxy bias model, computed from the inferred $\beta^\star$ following~\cref{eq:S8_from_beta}. The \ac{TFR}, \ac{SDSS} \ac{FP}, and \PP\ samples are in mutual agreement, and all except for the 6dF \ac{FP} sample agree well with the \textit{Planck} measurement ($0.832 \pm 0.013$;~\citealt{Planck_2020_cosmo}).}
    \label{fig:S8_1D_posterior}
\end{figure*}

We now present the $S_8$ measurements for the \ac{TFR}, \ac{FP}, and \acp{SN} samples.
We first analyse each catalogue independently and then perform a joint inference.
We convert the inferred velocity scaling parameter $\beta^\star$ into $S_8$ using
\begin{equation}\label{eq:S8_from_beta}
    S_8 = \sigma_{8}^{\rm L}\!\left(\sigma_8^{\rm NL} = \frac{\beta^\star\,\sigma_8^g}{\Om^{0.55}}\right) \sqrt{\frac{\Om}{0.3}},
\end{equation}
which we described in~\cref{sec:local_universe}, adopting the fiducial cosmology stated there. Varying $\Om$ to $0.25$ or $0.35$ changes the inferred $S_8$ by less than $1\sigma$.

\begin{figure*}
    \centering
    \includegraphics[width=1.0\textwidth]{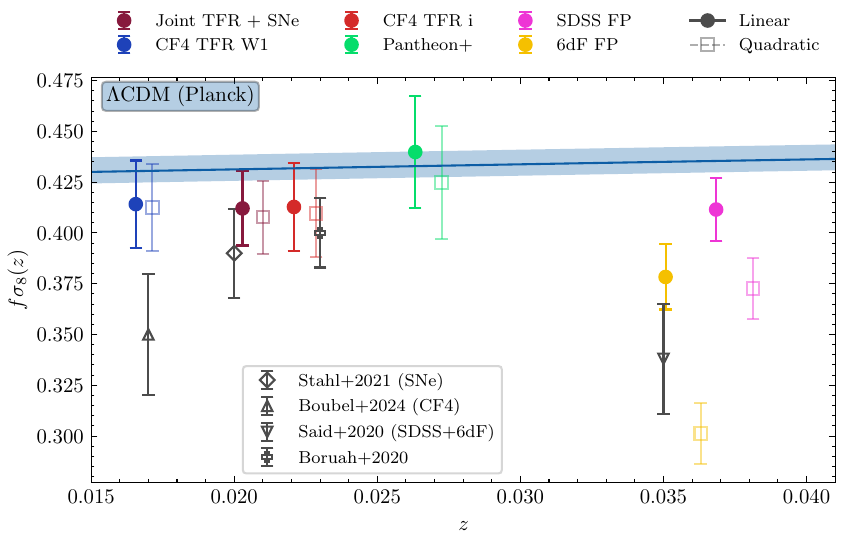}
    \caption{Comparison of $f\sigma_8^{\rm L}(z)$. Both our and literature measurements~\citep{Boruah_2019,Said_2020,Stahl_2021,Boubel_2024} are plotted at the effective redshifts of the samples, defined as their mean redshifts. Error bars denote $1\sigma$ uncertainties, and the \textit{Planck} prediction is shown as a $1\sigma$ shaded band~\citep{Planck_2020_cosmo}.}
    \label{fig:fsigma8}
\end{figure*}

\Cref{fig:S8_1D_posterior} shows the one-dimensional posterior predictive distributions of $S_8$ under the fiducial linear-bias model.
The \ac{TFR} and \PP\ samples lie within $0.3$--$0.7\sigma$ of the \textit{Planck} TT, TE, EE+lowE+lensing value $S_8 = 0.832 \pm 0.013$~\citep{Planck_2020_cosmo}.
Switching to the quadratic-bias variant does not significantly change the result for the \ac{TFR} and \acp{SN} samples, and they remain within $\lesssim 0.9\sigma$ of \textit{Planck}.
The \ac{FP} samples are more model-dependent: under the linear-bias model \ac{SDSS} \ac{FP} remains close to \textit{Planck}, while 6dF \ac{FP} lies $2.9\sigma$ below it.
Both \ac{FP} samples shift to lower $S_8$ under the quadratic-bias variant, showing sensitivity to the inhomogeneous Malmquist bias treatment.
Particularly in the \ac{FP} samples, the statistical uncertainty on $S_8$ is dominated by the uncertainty in $\sigma_8^g$, which propagates directly into $S_8$~\citep{Carrick_2015}.
Even with a perfectly measured $\beta^\star$, the uncertainty in $S_8$ would be $0.030$, which the two \ac{FP} samples approach.
Combining the two \ac{TFR} samples and \PP\ yields $S_8 = 0.798 \pm 0.035$ under the linear-bias model and $S_8 = 0.792 \pm 0.035$ under the quadratic-bias variant, with the uncertainty dominated by $\sigma_8^g$ in both cases.
We exclude the \ac{FP} samples from the joint inference.
Were we to include the \ac{SDSS} sample, it would dominate the $\beta^\star$ constraint because of its sample size, whereas the 6dF sample is in tension with the other datasets and thus we quote it separately (quantified in~\cref{tab:S8_tension}).
Finally, in~\cref{fig:fsigma8} we present constraints on $f\sigma_8^{\rm L}$ to enable comparison with other low-redshift probes.

In~\cref{tab:b1_beta_S8} we list the inferred values of the galaxy bias $b_1$ and $b_2$, velocity scaling $\beta^\star$, $f\sigma_8^{\rm L}$, and $S_8$ for both the fiducial linear-bias model and the quadratic-bias variant.
The linear bias $b_1$ varies significantly between samples and is always greater than unity, with the two \ac{FP} samples exhibiting the largest values, as expected for predominantly early-type, dispersion-supported ellipticals that occupy the densest environments and thus cluster more strongly; \PP\ shows an intermediate bias.
The same ordering carries over to the quadratic bias $b_2$: both \ac{TFR} samples prefer values consistent with zero, \PP\ a modest $b_2 = 0.36 \pm 0.09$, and the \ac{FP} samples sizeable values of $b_2 = 1.54 \pm 0.06$ (\ac{SDSS}) and $0.86 \pm 0.04$ (6dF), enhancing the non-linear response to high-contrast regions.
Relaxing the $b_2 > 0$ prior to permit negative values leaves both \ac{TFR} posteriors centred on zero rather than pressed against the boundary, with $S_8$ unchanged, while the \ac{FP} samples retain their large positive $b_2$; the quadratic-bias conclusions are therefore not driven by the prior.

After allowing for the different galaxy biases, the inferred values of $\beta^\star$ from the linear-bias model are in good agreement across \ac{CF4} W1, \ac{CF4} $i$, \PP, and \ac{SDSS} \ac{FP}, with only 6dF \ac{FP} preferring a lower value.
To quantify the internal agreement between our peculiar-velocity samples, we compute the pairwise tension on $S_8$ between each pair of samples using the suspiciousness statistic of~\citet{SPTSZ_clusters}\footnote{\url{https://github.com/SebastianBocquet/PosteriorAgreement}}, shown in~\cref{tab:S8_tension}; this is an internal consistency check among the peculiar-velocity datasets and does not involve \textit{Planck}.
Under linear bias, the largest pairwise tension is $1.9\sigma$, between 6dF \ac{FP} and \PP; the remaining pairs involving 6dF \ac{FP} sit between $1.2\sigma$ and $1.5\sigma$, and every pair not involving 6dF \ac{FP} lies below $1\sigma$.

The \ac{TFR} and \acp{SN} samples yield $\beta^\star$ values that are not significantly sensitive to the additional flexibility of the quadratic-bias model, whereas both \ac{FP} samples shift to noticeably lower values.
The effect is most pronounced for the 6dF \ac{FP} sample, whose $f\sigma_8$ drops to $0.301 \pm 0.015$, placing it in $7.8\sigma$ tension with \textit{Planck} and in $3.8$--$4.3\sigma$ tension with the \ac{TFR} and \acp{SN} samples (\cref{tab:S8_tension}).
As a further test, we add a cubic term to the inhomogeneous Malmquist bias model, but it is driven to zero in both \ac{FP} samples and so leaves the $\beta^\star$ and $S_8$ posteriors unaffected.

The quadratic extension is not self-consistent at the field level: it introduces a non-linear inhomogeneous Malmquist bias term while the~\citetalias{Carrick_2015} velocity field still derives from a linear-theory mapping of a linearly-biased \TWOMPP\ density field.
We therefore consider~\cref{eq:quadratic_bias} as a sensitivity test of the inferred $\beta^\star$ to the inhomogeneous Malmquist bias parametrisation.
Stability under this test is a necessary condition for the result to be modelling-independent at the precision quoted; instability flags that further work, ideally a non-linear forward reconstruction of the local density field such as \acl{BORG}~\citep{Jasche_2019,ManticoreLocal}, is needed.
The shift in $S_8$ between the linear and quadratic models is $0.08$ for \ac{SDSS} \ac{FP} and $0.15$ for 6dF \ac{FP}, both substantially exceeding the ${\sim} 0.03$ statistical uncertainty set by $\sigma_8^g$.
We accordingly do not regard the \ac{FP} $S_8$ as reliable at the precision otherwise quoted.
We cannot, with the present data and reconstruction, distinguish whether this shift reflects an unmodelled systematic of the \ac{FP} samples or a genuine non-linear contribution that the \ac{TFR} and \acp{SN} tracers, probing lower-density environments, do not detect; resolving this requires a non-linear forward model of the local density field along the lines of \ac{BORG}.
The \ac{FP} samples preferentially populate the peaks of the density field, where the linear-bias and linear-theory assumptions underlying the~\citetalias{Carrick_2015} reconstruction are most likely to break down.

Within the \ac{FP} class, the \ac{SDSS} and 6dF samples are consistent within $1.5\sigma$ under linear bias but driven into $3.3\sigma$ tension under the quadratic-bias variant, despite covering similar redshift ranges.
The two \ac{TFR} samples, by contrast, agree well despite differing sky coverage, disfavouring spatial variation in the reconstruction quality~\citep[a related question was examined by][in the context of masking]{Hollinger_2021} as the dominant cause.
We therefore adopt the \ac{TFR} and \acp{SN} as our fiducial tracers under the linear-bias model, and note that revisiting this comparison with the DESI \ac{FP} sample would be informative~\citep{Said_FPDESI}.

\begin{table*}
    \centering
    \setlength{\tabcolsep}{4pt}
    \begin{tabular}{lcccccc}
    Sample        & $z_{\rm eff}$ & $b_1$              & $b_2$              & $\beta^\star$            & $S_8$              & $f\sigma_8^{\rm L}$ \\
    \hline
    \multicolumn{7}{c}{\textit{Linear inhomogeneous Malmquist bias (fiducial)}} \\
    \hline
    CF4 TFR W1    & $0.017$ & $1.294 \pm 0.023$  & $-$  & $0.464 \pm 0.019$  & $0.802 \pm 0.042$  & $0.414 \pm 0.021$ \\
    CF4 TFR $i$   & $0.022$ & $1.213 \pm 0.018$  & $-$  & $0.462 \pm 0.020$  & $0.799 \pm 0.042$  & $0.413 \pm 0.022$ \\
    Pantheon+     & $0.026$ & $1.537 \pm 0.068$  & $-$  & $0.496 \pm 0.029$  & $0.851 \pm 0.053$  & $0.440 \pm 0.027$ \\
    SDSS FP       & $0.037$ & $2.054 \pm 0.030$  & $-$  & $0.460 \pm 0.006$  & $0.797 \pm 0.030$  & $0.412 \pm 0.016$ \\
    6dF FP        & $0.035$ & $1.979 \pm 0.024$  & $-$  & $0.419 \pm 0.010$  & $0.733 \pm 0.031$  & $0.379 \pm 0.016$ \\
    \textbf{Joint TFR + SNe}     & $-$ & $-$ & $-$ & $\mathbf{0.461 \pm 0.013}$ & $\mathbf{0.798 \pm 0.035}$ & $\mathbf{0.412 \pm 0.018}$ \\
    \hline
    \multicolumn{7}{c}{\textit{Quadratic inhomogeneous Malmquist bias}} \\
    \hline
    CF4 TFR W1    & $0.017$ & $1.294 \pm 0.023$  & $-0.002 \pm 0.017$  & $0.462 \pm 0.019$  & $0.799 \pm 0.041$  & $0.413 \pm 0.021$ \\
    CF4 TFR $i$   & $0.022$ & $1.214 \pm 0.018$  & $0.004 \pm 0.015$  & $0.458 \pm 0.020$  & $0.793 \pm 0.042$  & $0.410 \pm 0.022$ \\
    Pantheon+     & $0.026$ & $1.581 \pm 0.070$  & $0.364 \pm 0.089$  & $0.477 \pm 0.029$  & $0.822 \pm 0.054$  & $0.425 \pm 0.028$ \\
    SDSS FP       & $0.037$ & $2.318 \pm 0.033$  & $1.544 \pm 0.060$  & $0.412 \pm 0.007$  & $0.721 \pm 0.029$  & $0.373 \pm 0.015$ \\
    6dF FP        & $0.035$ & $2.059 \pm 0.023$  & $0.861 \pm 0.040$  & $0.326 \pm 0.012$  & $0.583 \pm 0.029$  & $0.301 \pm 0.015$ \\
    Joint TFR + SNe                & $-$ & $-$ & $-$ & $0.457 \pm 0.013$ & $0.792 \pm 0.035$ & $0.409 \pm 0.018$ \\
    \hline
    \end{tabular}
    \caption{The effective redshift $z_{\rm eff}$, galaxy biases $b_1$ and $b_2$, velocity scaling parameter $\beta^\star$, $f\sigma_8^{\rm L}$, and thus $S_8$ for the two \ac{TFR} samples (\ac{CF4} W1 and \ac{CF4} $i$), the two \ac{FP} samples (\ac{SDSS} and 6dF), and \PP, under the fiducial linear inhomogeneous Malmquist bias model and the quadratic-bias robustness variant of~\cref{eq:quadratic_bias}. Bold marks the fiducial joint constraint, which combines the \ac{TFR} and \PP\ samples. The effective redshift is computed as the mean \ac{CMB} redshift of each sample.}
    \label{tab:b1_beta_S8}
\end{table*}

\begin{table}
    \centering
    \setlength{\tabcolsep}{4pt}
    \begin{tabular}{lccccc}
    \hline
               & CF4 W1 & CF4 $i$ & \PP & SDSS FP & 6dF FP \\
    \hline
    \multicolumn{6}{c}{\textit{Linear inhomogeneous Malmquist bias (fiducial)}} \\
    \hline
    CF4 W1     & $-$    &         &      &         &        \\
    CF4 $i$    & 0.04   & $-$     &      &         &        \\
    \PP        & 0.73   & 0.76    & $-$  &         &        \\
    SDSS FP    & 0.08   & 0.09    & 0.88 & $-$     &        \\
    6dF FP     & 1.29   & 1.23    & 1.88 & 1.47    & $-$    \\
    \hline
    \multicolumn{6}{c}{\textit{Quadratic inhomogeneous Malmquist bias}} \\
    \hline
    CF4 W1     & $-$    &         &      &         &        \\
    CF4 $i$    & 0.14   & $-$     &      &         &        \\
    \PP        & 0.29   & 0.42    & $-$  &         &        \\
    SDSS FP    & 1.51   & 1.35    & 1.67 & $-$     &        \\
    6dF FP     & 4.28   & 4.08    & 3.78 & 3.28    & $-$    \\
    \hline
    \end{tabular}
    \caption{Pairwise tension statistic, in units of $\sigma$, between the $S_8$ posteriors from each pair of datasets, computed using the suspiciousness statistic of~\protect\citet{SPTSZ_clusters} under the fiducial linear bias (top) and the quadratic-bias variant (bottom).}
    \label{tab:S8_tension}
\end{table}

\begin{figure}
    \centering
    \includegraphics[width=\columnwidth]{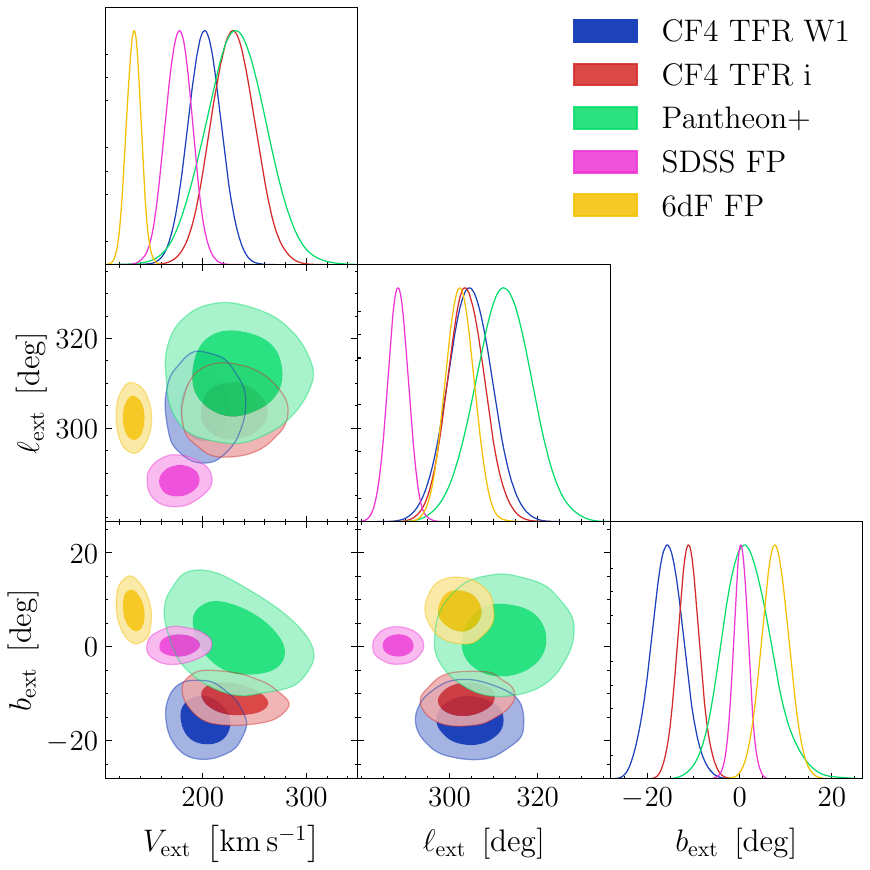}
    \caption{Posterior distribution of the external flow vector $\Vext$, shown in terms of its magnitude $V_{\rm ext}$ and direction $(\ell_{\rm ext},\,b_{\rm ext})$ in Galactic coordinates. The inferred magnitudes are broadly consistent across samples, whereas the directions show mild discrepancies: the two \ac{TFR} samples agree with each other but differ from the two \ac{FP} samples, which themselves are not mutually consistent. The contours are $1$ and $2\sigma$.}
    \label{fig:Vext_comparison}
\end{figure}

Finally, as part of our flow model we infer $\Vext$, the external flow vector, and compare its magnitude and direction in~\cref{fig:Vext_comparison}.
The inferred magnitudes are in good overall agreement: most samples prefer ${\sim} 200 \pm 20~\kmsec$, with the exception of the 6dF \ac{FP} sample, which yields a smaller value of $140 \pm 7~\kmsec$.
Larger differences arise in the inferred directions.
The two \ac{CF4} \ac{TFR} samples are consistent with each other, while the \ac{SDSS} and 6dF \ac{FP} samples point in different directions, separated by $20 \pm 4^\circ$.
This discrepancy may reflect their different sky coverage and/or survey depth.
By contrast,~\citet{Said_2020} analysed the \ac{SDSS} and 6dF \ac{FP} samples and found good agreement in the direction of $\Vext$, although we do not use the exact same data, with the differences summarised in~\cref{sec:comparison_to_literature}.
The \PP\ dipole is less well constrained than the others and falls between the 6dF \ac{FP} and the two \ac{TFR} samples, but differs from the \ac{SDSS} \ac{FP} direction by $27 \pm 7^\circ$.
When combining the datasets in a joint inference, we assume a common $\Vext$ vector across all samples, even though, as shown in~\cref{fig:Vext_comparison}, the \ac{TFR} samples and \PP\ prefer slightly different $\Vext$.
Allowing a separate $\Vext$ per sample leaves the joint $\beta^\star$ essentially unchanged.

%%%%%%%%%%%%%%%%%%%%%%%%%%%%%%%%%%%%%%%%%%%%%%%%%%%%%%%%%%%%%%%%%%%%%%%%%%%%%%%
%                             Discussion                                      %
%%%%%%%%%%%%%%%%%%%%%%%%%%%%%%%%%%%%%%%%%%%%%%%%%%%%%%%%%%%%%%%%%%%%%%%%%%%%%%%

\section{Comparison to literature}\label{sec:comparison_to_literature}

\begin{figure}
    \centering
    \includegraphics[width=\columnwidth]{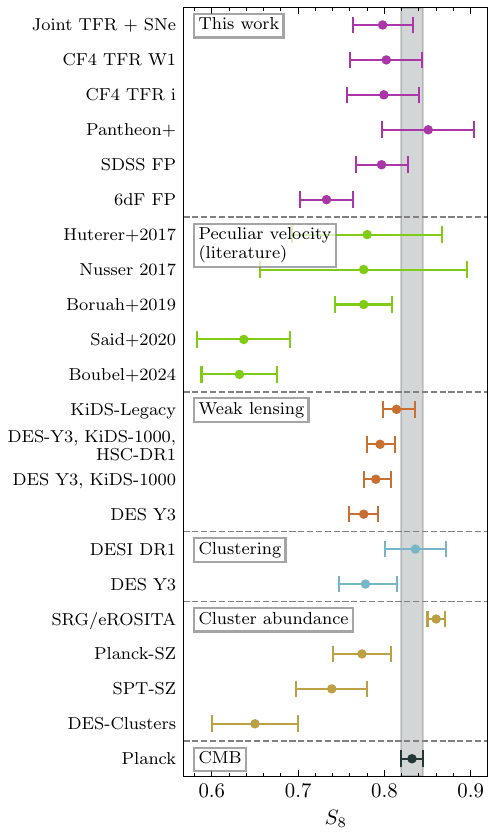}
    \caption{The $S_8$ parameter inferred by calibrating the~\protect\citet{Carrick_2015} linear field against the peculiar velocity samples. We compare to literature results using peculiar velocities~\protect\citep{Huterer_2017,Nusser_2017,Boruah_2019,Said_2020}, weak lensing~\protect\citep{Wright_2025,Carlos_2024,DES_KIDS,DES_Y3}, clustering~\protect\citep{DESI_DR1_fullshape,DESY3_clustering}, cluster abundance~\protect\citep{Ghirardini_2024,SPTSZ_clusters,Planck_clusters,DES_clusters} and \textit{Planck} TT, TE, EE+lowE+lensing~\protect\citep{Planck_2020_cosmo}. The errors are $1\sigma$.}
    \label{fig:S8_comparison}
\end{figure}

Using the linear \TWOMPP\ density and velocity field of~\citetalias{Carrick_2015}, we find that the fiducial combination of \ac{TFR} and \acp{SN} yields $S_8$ in agreement with \textit{Planck}, while the \ac{FP} constraints are only mildly discrepant under the linear-bias model and become unreliable under the quadratic-bias sensitivity test.
In~\cref{fig:S8_comparison} we quantify this comparison, showing both our inferred values of $S_8$ and literature measurements.
Given the measurement precision, we cannot distinguish between the \ac{CMB} \textit{Planck} result and weak lensing studies~\citep{Wright_2025,Carlos_2024,DES_KIDS,DES_Y3} in the context of the $S_8$ tension, and correspondingly we find no appreciable discrepancy with either.
We also find agreement with clustering analyses~\citep{DESI_DR1_fullshape,DESY3_clustering}, whose constraints on $S_8$ are less precise than our fiducial joint measurement.

Our measurements are consistent with several previous $S_8$ measurements from peculiar velocities~\citep{Huterer_2017,Nusser_2017,Boruah_2019}, the last of which also employed the~\citetalias{Carrick_2015} density and velocity fields.
However, we find large disagreement with~\citet{Said_2020} and~\citet{Boubel_2024}, who likewise inferred $S_8$ by calibrating the $\beta^\star$ parameter of~\citetalias{Carrick_2015}.
The former used the 6dF and \ac{SDSS} \ac{FP} samples, while the latter analysed the \ac{CF4} W1 and \ac{CF4} $i$ samples.

We attribute the discrepancy primarily to methodological differences.
First,~\citeauthor{Boubel_2024} adopt a ``partial'' forward model: rather than sampling distances and querying the (real-space)~\citetalias{Carrick_2015} field at those distances, they map that field to redshift space following the approach of~\citet{Carr_2022}, assuming some fiducial $\Vext$ and $\beta^\star$.
They then query the redshift-space field at the observed redshift of each tracer to obtain the \ac{LOS} peculiar velocity and apply an approximate correction to replace the fiducial $\Vext$ and $\beta^\star$ with sampled values.
Second, their flow model is formulated in a way that dispenses with inhomogeneous Malmquist bias, and thus omits the tracer density field altogether.

However, this formulation is inconsistent with Bayesian methodology.
Within their Eq.~2,~\citeauthor{Boubel_2024} consider the joint distribution of the observed and cosmological redshifts, which they factorise as
\begin{equation}
    p(\zobs,\,\zcosmo \mid \ldots)
    =
    p(\zcosmo | \zobs,\,\ldots) p(\zobs\mid \ldots),
\end{equation}
where they say that $p(\zcosmo | \zobs,\,\ldots)$ is a Gaussian distribution centred at $\zobs$ with a standard deviation $\sigma_v$ and place a delta-function ``prior'' on $\zobs$ (because it is known).
However, in a Bayesian model one must specify the likelihood of observed data given some model parameters, and assign a prior distribution to the model parameters.
There cannot be a prior distribution of the observed data.
Instead, the correct treatment is to specify the likelihood of $\zobs$ given the model parameters (which is Gaussian), together with an appropriate prior on $\zcosmo$.
Since $\zcosmo$ is related to distance, its prior would then be subject to inhomogeneous Malmquist bias, contrary to what~\citeauthor{Boubel_2024} claim.

\begin{figure}
    \centering
    \includegraphics[width=\columnwidth]{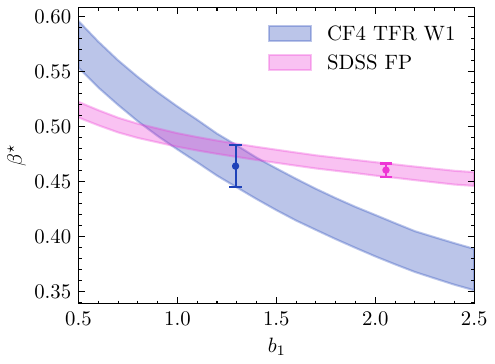}
    \caption{Inferred $\beta^\star$ as a function of the fixed linear bias $b_1$ for the \ac{CF4} W1 and \ac{SDSS} \ac{FP} samples, exposing the strong $b_1$--$\beta^\star$ degeneracy that motivates sampling both parameters jointly in the fiducial analysis.
    }
    \label{fig:b1_beta}
\end{figure}

We next compare with~\citeauthor{Said_2020}, who analysed (nearly) the same two \ac{FP} samples as we do and measured $f\sigma_8 = 0.338 \pm 0.027$, well below \textit{Planck}.
Under our fiducial linear-bias model the \ac{FP} samples are not severely discrepant: the \ac{SDSS} \ac{FP} sample yields $f\sigma_8 = 0.412 \pm 0.016$ and 6dF \ac{FP} $0.379 \pm 0.016$, broadly consistent with our \ac{TFR} and \PP\ values and only mildly below \textit{Planck}.
The strongly low values appear once the inhomogeneous Malmquist bias is extended to a quadratic-bias model, under which both \ac{FP} samples migrate further from \textit{Planck} (in particular 6dF \ac{FP} reaches $f\sigma_8 = 0.301 \pm 0.015$); crucially, the \ac{FP} data themselves prefer this extension over the linear-bias model.
\citeauthor{Said_2020}, by adopting the prior $p(r) \propto r^2 \left[1+\delta_g(r)\right]$, implicitly fix $b_1 = 1$ and do not model the galaxy bias at all.
This assumption alone cannot explain their low $f\sigma_8$: as~\cref{fig:b1_beta} shows, fixing $b_1 = 1$ in our \ac{SDSS} \ac{FP} pipeline drives $\beta^\star$ up to ${\sim} 0.51$, well above its free-bias value, so the linear-bias choice acts in the opposite direction to the deficit they report.
As in the case of~\citeauthor{Boubel_2024}, one difference arises from the treatment of inhomogeneous Malmquist bias.
Unlike~\citeauthor{Boubel_2024}, they work consistently in real space.
However, in their Eq.~23 they adopt a prior on the source distance $p(r) \propto r^2 \left[1+\delta_g(r)\right]$, where $\delta_g$ is the luminosity density contrast of the~\citetalias{Carrick_2015} field.
This corresponds to the assumption $b_1 = 1$ in our formulation (see Eq.~\ref{eq:linear_bias}).
As shown in~\cref{tab:b1_beta_S8}, all samples prefer values of $b_1$ significantly different from unity, indicating that these galaxy samples are not unbiased tracers of the \TWOMPP\ galaxy field, as implicitly assumed by~\citeauthor{Said_2020}.
Selection effects are another significant difference.
\citeauthor{Said_2020} model selection by taking the per-galaxy likelihood to the power of $1/S_n$, where $S_n$ is the fraction of the survey volume over which the $n$\textsuperscript{th} galaxy could be observed.
From a Bayesian perspective, the $1/S_n$ likelihood weighting lacks clear justification from first principles.
Nevertheless, we likewise resort to approximate selection modelling through the distance prior inspired by~\citet{Lavaux_Virbius} given in~\cref{eq:empirical_prior_distance}, but our approach infers the selection parameters during inference rather than fixing them a priori.
Another difference is that in their Eq.~24 they formulate the likelihood in terms of the effective angular size itself rather than its logarithm, as we do.
For galaxies of the same physical size, the angular size scales inversely with distance.
Thus, assuming a constant intrinsic scatter in angular size (rather than its logarithm) implies that the scatter represents different physical scales for nearby versus distant galaxies.
By formulating the likelihood in logarithmic angular size, the intrinsic scatter corresponds to a multiplicative uncertainty that is scale-invariant with distance, which is more physically motivated.

\Cref{fig:b1_beta} shows the inferred $\beta^\star$ from the \ac{CF4} W1 and \ac{SDSS} \ac{FP} samples when $b_1$ is fixed during inference.
We test values of $b_1$ uniformly spaced between 0.5 and 2.5.
The figure highlights the degeneracy between $b_1$ and $\beta^\star$: fixing $b_1$ below the model-preferred range yields artificially high $\beta^\star$.
This behaviour is expected: larger $b_1$ values increase the probability of inferred distances placing galaxies in overdense regions where peculiar velocities are higher, thereby reducing the $\beta^\star$ required to match the true velocities.

The amplitude of the residual offset between our \ac{FP} $\beta^\star$ and that of~\citeauthor{Said_2020} likely reflects the remaining methodological differences, for example modelling the intrinsic scatter in angular size rather than in its logarithm, or differences in the adopted samples (we use the \ac{SDSS} \ac{FP} compilation of~\citealt{Howlett_2022}, whereas \citeauthor{Said_2020} employed an earlier version of the \ac{SDSS} \ac{FP} sample).
In comparison with~\citeauthor{Boubel_2024}, it is likewise uncertain how their treatment affects $\beta^\star$.
By sampling the cosmological redshift directly from the observed redshift, their model bypasses a treatment of inhomogeneous Malmquist bias: the distance prior is effectively replaced by the empirical number density of the sample.

%%%%%%%%%%%%%%%%%%%%%%%%%%%%%%%%%%%%%%%%%%%%%%%%%%%%%%%%%%%%%%%%%%%%%%%%%%%%%%%
%                             Conclusion                                      %
%%%%%%%%%%%%%%%%%%%%%%%%%%%%%%%%%%%%%%%%%%%%%%%%%%%%%%%%%%%%%%%%%%%%%%%%%%%%%%%

\section{Conclusion}\label{sec:conclusion}

We have shown that the \ac{TFR} (the \ac{CF4} \ac{WISE} W1 and \ac{SDSS} $i$ samples) and \acp{SN} (\PP) direct-distance tracers provide competitive and mutually consistent measurements of $S_8$, while the \ac{FP} samples (\ac{SDSS} and 6dF) yield constraints that strongly depend on the inhomogeneous Malmquist bias treatment, calling the validity of their cosmological constraints into question.
The former are also consistent with the \textit{Planck} \ac{CMB} result, weak lensing, and galaxy clustering.
Under the quadratic-bias variant the \ac{FP} samples shift to still lower values, reaching ${\sim} 3.5\sigma$ tension with \textit{Planck} for \ac{SDSS} \ac{FP} and ${\sim} 7.8\sigma$ for 6dF \ac{FP}.
Nevertheless, our fiducial joint constraint, combining the \ac{TFR} and \PP\ samples, is $S_8 = 0.798 \pm 0.035$ ($f\sigma_8 = 0.412 \pm 0.018$), in agreement with \textit{Planck}, with the uncertainty dominated by the variance of the \TWOMPP\ galaxy field ($\sigma_8^g$).

\citet{Hollinger_2024} used mock \TWOMPP\ realisations together with mock \ac{TFR} catalogues to argue that the sample variance in $f\sigma_{8,\mathrm{NL}} = \beta^\star \sigma_8^g$ is at most five per cent.
Propagating this additional uncertainty to $S_8$ yields $S_8 = 0.798 \pm 0.053$.
This estimate is overly conservative, since it was derived for significantly smaller catalogues than those used here.
The assumption of linear bias in both the~\citetalias{Carrick_2015} reconstruction and our inhomogeneous Malmquist bias modelling is a limitation.
However, the \TWOMPP\ density field is smoothed on scales of $4~\Mpch$, and the velocity field is sensitive to even larger scales due to the $1/k^2$ weighting in the velocity power spectrum~\citep{Peebles1980}.
Moreover,~\citet{Hollinger_2024} tested the linear bias assumption in the~\citetalias{Carrick_2015} field using simulations and found it to be adequate.
We further explored a quadratic extension of the inhomogeneous Malmquist bias: the two \ac{TFR} samples and \PP\ are stable under this extension, whereas both \ac{FP} samples are not, and we accordingly do not regard the present \ac{FP} constraints as reliable.

More sophisticated non-linear approaches will become feasible with forward-modelling methods such as \acl{BORG}~\citep{Jasche_2019,ManticoreLocal}. When converting $\beta^\star \sigma_8^g$ to $S_8$, we assume $\Om = 0.3111$.
By definition of $S_8$ and using $\sigma_8^{\rm NL} = \beta^\star \sigma_8^g / \Om^{0.55}$, the inferred $S_8$ depends only weakly on $\Om$.
If we had instead assumed $\Om = 0.25$ or $0.35$, the resulting $S_8$ would shift by $-3.8$ or $+2.1$ per cent ($-1.0\sigma$ or $+0.6\sigma$), respectively.
Because of the limited redshift range of our samples, we cannot constrain $\Om$ from these data alone and therefore adopt the \textit{Planck} value.
Relatedly,~\citet{Blake_2024} emphasised that analyses which calibrate the velocity field may underestimate uncertainty in $\beta^\star$ if the residual peculiar velocity covariance is not fully modelled.
In our analysis we assume a diagonal residual redshift covariance.
Our results agree with several previous peculiar velocity inferences of the growth rate, with more precise constraints.
The \ac{TFR} and \PP\ samples are in tension with the lower values reported by~\citet{Said_2020} and~\citet{Boubel_2024}; our own \ac{FP} samples reproduce a low $f\sigma_8$ qualitatively, but only when the inhomogeneous Malmquist bias is extended to a quadratic-bias model, while~\citet{Said_2020} fix the linear bias to unity and do not model it at all.
We argue this convergence likely reflects a shared limitation of linear-theory \ac{FP} analyses in dense environments rather than a robust late-time growth deficit, and we identify methodological issues in these works that plausibly set the size of the residual offset.

Recent indications of dynamical dark energy (e.g.~\citealt{DESI_DR2_BAO_2025}) make a low-redshift measurement of $f\sigma_8$ particularly timely.
Non-minimally coupled or Galileon dark energy models predict an enhanced $f\sigma_8$ at $z = 0$ relative to \ac{LCDM}~\citep{Wolf_2025}.
Our results can therefore provide constraints on this class of models.
Similarly,
these results are highly relevant in light of the $S_8$ tension, which has been considered a tension between early- and late-Universe measurements of $S_8$.
This tension has been driven primarily by weak lensing surveys, which have preferred values lower than \textit{Planck}~\citep{Heymans_2013,Asgari_2021,Li_2023,Amon:2022,Secco_2022,Preston_2023,Li_2023_B,Dalal_2023,DES_KIDS}; only~\citet{Jee_2016} reported $S_8$ slightly higher than \textit{Planck}.
More recently, however, the Kilo-Degree Survey (KiDS) reported $S_8 = 0.815^{+0.016}_{-0.021}$~\citep{Wright_2025}, in good agreement with \textit{Planck}, attributing the shift to improved redshift estimation and calibration.
By contrast, the Dark Energy Survey Year~3 recently reported $S_8 = 0.780 \pm 0.015$~\citep{Gomes_2025}, which remains in mild tension with \textit{Planck}.
We show that, when analysed with a Bayesian hierarchical model, the \ac{TFR} and \acp{SN} samples yield $S_8$ in agreement with \textit{Planck} rather than preferring lower values, motivating further scrutiny of \ac{FP}-specific systematics. Our peculiar velocity results therefore favour concordance with the \ac{CMB} rather than the early-versus-late dichotomy.

\section{Data availability}

The~\citet{Carrick_2015} reconstruction is available at~\href{https://cosmicflows.iap.fr}{cosmicflows.iap.fr}.
The public \ac{CF4} data (both \ac{TFR} and \ac{SDSS} \ac{FP}) are available at~\href{https://edd.ifa.hawaii.edu/dfirst.php}{edd.ifa.hawaii.edu}.
The 6dF \ac{FP} sample is available at~\href{https://cdsarc.cds.unistra.fr/viz-bin/cat/J/MNRAS/443/1231#/browse}{CDS archive}.
The public \PP\ data release is available at~\href{https://github.com/PantheonPlusSH0ES/DataRelease}{github.com/PantheonPlusSH0ES}.
The code and all other data will be made available on reasonable request to the authors.

\section*{Acknowledgements}

We thank Harry Desmond, Julien Devriendt, Pedro G. Ferreira, Mike Hudson, Guilhem Lavaux, and Adrianne Slyz for useful inputs and discussions. We thank Anthony Carr for providing a version of the~\PP~covariance matrix with the peculiar velocity contributions removed. The authors would like to acknowledge the use of the University of Oxford Advanced Research Computing (ARC) facility in carrying out this work.\footnote{\url{https://doi.org/10.5281/zenodo.22558}} RS acknowledges financial support from STFC Grant No. ST/X508664/1, the Snell Exhibition of Balliol College, Oxford.

\bibliographystyle{mnras}
\bibliography{ref}

\bsp
\label{lastpage}
\end{document}